\documentclass[12pt]{article}

\usepackage{amsmath,amsfonts,amssymb,latexsym,cite,bm,color} 
\usepackage{slashed} 
\usepackage{relsize} 
\usepackage{graphicx} 

\usepackage{hyperref}  

\usepackage{prodint}

\makeatletter
\def\@prodint#1{\vcenter{\hbox{#1${\pi}$}}}  
\def\prodint{\mathop{
  \mathchoice{\@prodint\Huge}{\@prodint\Large}
	{\@prodint\large}{\@prodint\normalsize}}}
\makeatother 

\newcounter{thMM}
\newcounter{leMM}
\newcounter{deFF}
\newcounter{exMP}
\title{
 \small\bf 
SCHR\"ODINGER FUNCTIONAL OF A QUANTUM SCALAR FIELD IN \\
\vspace*{0.3em} 
STATIC SPACE-TIMES 
FROM PRECANONICAL QUANTIZATION 
\\
}
\author{ \normalsize
Igor V. Kanatchikov
\\ 
\small\it School of Physics and Astronomy \\ 
\small\it University of St Andrews, St Andrews KY16 9SS, UK \\
[0.5ex]
\small\it Quantum Information Center in Gda\'nsk (KCIK),\\ 
\small\it 81-831 Sopot, Poland  
}

\date{}

  \catcode `\@=11
 
\def\section{\@startsection {section}{1}{\z@}{-3.5ex plus -1ex minus
     -.2ex}{2.3ex plus .2ex}{\normalsize\bf}}
\def\subsection{\@startsection{subsection}{2}{\z@}{-3.25ex plus -1ex minus
 -.2ex}{1.5ex plus .2ex}{\normalsize\bf}}

 \def\thebibliography#1{\section*{References\markboth
  {REFERENCES}{REFERENCES}}\list
  {[\arabic{enumi}]}{\settowidth\labelwidth{[#1]}\leftmargin\labelwidth
  \advance\leftmargin\labelsep
  \usecounter{enumi}}
  \def\newblock{\hskip .11em plus .33em minus -.07em}
  \sloppy
  \sfcode`\.=1000\relax} 
 \let\endthebibliography=\endlist
 
\catcode `\@=12

\begin{document}


\maketitle
\begin{abstract}
\noindent 
{\footnotesize 
The functional Schr\"odinger representation of a scalar field on an $n$-dimensional 
static space-time background is argued to be a singular limiting case of the hypercomplex quantum theory of the same system obtained by the precanonical quantization based on the space-time symmetric De Donder-Weyl Hamiltonian theory. The functional Schr\"odinger representation emerges from the precanonical quantization  when the ultraviolet parameter $\varkappa$ introduced by precanonical quantization is replaced by $\underline{\gamma}{}_0\delta^\mathrm{inv}(\mathbf{0})$, where $\underline{\gamma}{}_0$ is the time-like tangent space Dirac matrix and $\delta^\mathrm{inv}(\mathbf{0})$ is an invariant spatial $(n-1)$-dimensional Dirac's delta function whose regularized value at $\mathbf{x}=\mathbf{0}$  is identified with the cutoff of the volume of the momentum space. In this limiting case, the Schr\"odinger wave functional is expressed as the trace of the product integral  of Clifford-algebra-valued precanonical wave  functions restricted to a certain field configuration and the canonical functional derivative Schr\"odinger equation is derived from the manifestly covariant Dirac-like precanonical Schr\"odinger equation which is independent of a choice of a codimension-one foliation. 
 }

\medskip


\end{abstract}

\newcommand{\jrefignre}{ 
\vspace*{-177mm}  
\hbox to 5.8truein{
\small to appear in Int.~J.~Geom.~Meth.~Mod.~Phys. {\bf 16} (2019) 1950017, \hfil 
\hbox to 1 truecm{ 
\hss \small arXiv:1810.09968v2
} 
\vspace*{-4mm}
}
\hbox to 5.8truein{\hspace*{-4pt}\hrulefill}
\vspace*{175mm}
 } 


\newcommand{\beq}{\begin{equation}}
\newcommand{\eeq}{\end{equation}}
\newcommand{\beqa}{\begin{eqnarray}}
\newcommand{\eeqa}{\end{eqnarray}}
\newcommand{\nn}{\nonumber}

\newcommand{\half}{\frac{1}{2}}

\newcommand{\xt}{\tilde{X}}

\newcommand{\uind}[2]{^{#1_1 \, ... \, #1_{#2}} }
\newcommand{\lind}[2]{_{#1_1 \, ... \, #1_{#2}} }
\newcommand{\com}[2]{[#1,#2]_{-}} 
\newcommand{\acom}[2]{[#1,#2]_{+}} 
\newcommand{\compm}[2]{[#1,#2]_{\pm}}

\newcommand{\lie}[1]{\pounds_{#1}}
\newcommand{\co}{\circ}
\newcommand{\sgn}[1]{(-1)^{#1}}
\newcommand{\lbr}[2]{ [ \hspace*{-1.5pt} [ #1 , #2 ] \hspace*{-1.5pt} ] }
\newcommand{\lbrpm}[2]{ [ \hspace*{-1.5pt} [ #1 , #2 ] \hspace*{-1.5pt}
 ]_{\pm} }
\newcommand{\lbrp}[2]{ [ \hspace*{-1.5pt} [ #1 , #2 ] \hspace*{-1.5pt} ]_+ }
\newcommand{\lbrm}[2]{ [ \hspace*{-1.5pt} [ #1 , #2 ] \hspace*{-1.5pt} ]_- }

\newcommand{\pbr}[2]{ \{ \hspace*{-2.2pt} [ #1 , #2\hspace*{1.4 pt} ] 
\hspace*{-2.3pt} \} }
\newcommand{\nbr}[2]{ [ \hspace*{-1.5pt} [ #1 , #2 \hspace*{0.pt} ] 
\hspace*{-1.3pt} ] }

\newcommand{\we}{\wedge}
\newcommand{\nbrpq}[2]{\nbr{\xxi{#1}{1}}{\xxi{#2}{2}}}
\newcommand{\lieni}[2]{$\pounds$${}_{\stackrel{#1}{X}_{#2}}$  }

\newcommand{\rbox}[2]{\raisebox{#1}{#2}}
\newcommand{\xx}[1]{\raisebox{1pt}{$\stackrel{#1}{X}$}}
\newcommand{\xxi}[2]{\raisebox{1pt}{$\stackrel{#1}{X}$$_{#2}$}}
\newcommand{\ff}[1]{\raisebox{1pt}{$\stackrel{#1}{F}$}}
\newcommand{\dd}[1]{\raisebox{1pt}{$\stackrel{#1}{D}$}}
\newcommand{\der}{\partial}
\newcommand{\oo}{$\Omega$}
\newcommand{\Om}{\Omega}
\newcommand{\om}{\omega}
\newcommand{\eps}{\epsilon}
\newcommand{\si}{\sigma} 
\newcommand{\Si}{\Sigma} 
\newcommand{\Lm}{\bigwedge^*}

\newcommand{\inn}{\hspace*{2pt}\raisebox{-1pt}{\rule{6pt}{.3pt}\hspace*
{0pt}\rule{.3pt}{8pt}\hspace*{3pt}}}
\newcommand{\sro}{Schr\"{o}dinger\ }
\newcommand{\vol}{\omega}
               \newcommand{\dvol}[1]{\der_{#1}\inn \vol}

\newcommand{\bd}{\mbox{\bf d}}
\newcommand{\bder}{\mbox{\bm $\der$}}
\newcommand{\bI}{\mbox{\bm $I$}}

\newcommand{\be}{\beta} 
\newcommand{\ga}{\gamma} 
\newcommand{\dlt}{\delta} 
\newcommand{\Ga}{\Gamma} 
\newcommand{\gmu}{\gamma^\mu}
\newcommand{\gnu}{\gamma^\nu}
 \newcommand{\ka}{\varkappa} 
 \newcommand{\la}{\lambda}
\newcommand{\hka}{\hbar \kappa}
\newcommand{\al}{\alpha}
\newcommand{\lapl}{\bigtriangleup}
\newcommand{\psib}{\overline{\psi}}
\newcommand{\Psib}{\overline{\Psi}}
\newcommand{\Phib}{\overline{\Phi}}
\newcommand{\derts}{\stackrel{\leftrightarrow}{\der}}
\newcommand{\what}[1]{\widehat{#1}}

\newcommand{\gmab}{\bar{\gamma}}
\newcommand{\gammat}{\; \tilde{\gamma}{}}

\newcommand{\bx}{{\bf x}}
\newcommand{\bk}{{\bf k}}
\newcommand{\bq}{{\bf q}}

\newcommand{\omk}{\omega_{\bf k}} 
\newcommand{\lpl}{\ell}
\newcommand{\zb}{\overline{z}} 

\newcommand{\dltb}{\bar \dlt}

\newcommand{\dv}{\mbox{\sf d}}

\newcommand{\deltt}{\bm{\dlt}}   

\newcommand{\Psbld}{\mathbf{\Psi}} 
\newcommand{\BPhi}{\mathbf{\Phi}}
\newcommand{\BH}{{\bf H}} 
\newcommand{\BS}{{\bf S}} 
\newcommand{\BN}{{\bf N}} 
\newcommand{\BXi}{{\bf \Xi}}

\newcommand{\rd}{\mathrm{d}}
\newcommand{\ri}{\mathrm{i}}
\newcommand{\Tr}{\mathrm{Tr}} 

\newcommand{\boldx}{{\bx}}

\medskip 

\section{Introduction}  

\newcommand{\fnc}{{\rm function}}
\newcommand{\equn}{{\rm equation}}
\newcommand{\ota}{{\rm obtain}}
\newcommand{\dwh}{{\rm DW Hamiltonian}}
\newcommand{\fd}{{\rm field}}
\newcommand{\con}{{\rm connection}}
\newcommand{\bewn}{{\rm between}}
\newcommand{\prc}{{\rm precanonical}}
 \newcommand{\vsh}{{\rm vanish}}
 \newcommand{\lmt}{{\rm limit}}
 \newcommand{\tmm}{{\rm term}}
 \newcommand{\cse}{{\rm case}}
 \newcommand{\Sch}{{\rm Schr\"{o}dinger}}
 \newcommand{\wnn}{{\rm when}}
 \newcommand{\ltr}{{\rm latter}}
 \newcommand{\rsd}{{\rm restricted}}
  \newcommand{\drv}{{\rm derivat}}
 \newcommand{\clf}{{\rm Clifford}}
\newcommand{\fst}{{\rm first}}
\newcommand{\rln}{{\rm relation}}
\newcommand{\dfrt}{{\rm different}}
\newcommand{\rsp}{{\rm respect}}
\newcommand{\by}{{\rm by}}
\newcommand{\spt}{{\rm space-time}}
\newcommand{\crsp}{{\rm correspond}}
\newcommand{\trt}{{\rm treat}}
\newcommand{\hlt}{{\rm Hamilton}} 
\newcommand{\and}{{\rm and}}
\newcommand{\qnt}{{\rm quant}} 
\newcommand{\myy}{{\rm we}}
\newcommand{\imls}{{\rm implies}}
\newcommand{\dnt}{{\rm denote}}
\newcommand{\gma}{\gamma}

The canonical \qnt{}ization of scalar \fd{} theory in curved \spt{} using the 
functional \Sch{} picture of QFT 
  \cite{fr,pi,se1,se2,cr1,cr2,kr}) 
leads to the description of the \crsp{}ing \qnt{}um \fd{} in \tmm{}s of the \Sch{} 
wave \fnc{}al $\Psbld([\phi(\boldx)],t)$  which satisfies the 
\fnc{}al \drv{}ive \Sch{} \equn{} 
\beq \label{fuse}
\ri\hbar \partial_t \Psbld = 
 \int \! d\boldx\, \sqrt{-g}
 \left (\frac{\hbar^2}{2}\ \frac{g_{00}}{g}\frac{\dlt^2}{\dlt \phi(\boldx)^2} 
- \frac{1}{2} g{}^{ij} \partial_i\phi(\bx)\partial_j\phi(\boldx) + V(\phi)
 \right) \Psbld , 
\eeq 
where $g$${}_{\mu\nu} =g$${}_{\mu\nu}(x)$ \dnt{}s the metric tensor, 
$g$ is the de\tmm{}inant of $g$${}_{\mu\nu}$,  \and{} $x^\mu = (t, x^i) = (t,\boldx)$ 
are \spt{} coordinates.  
In writing this \equn{}, one chooses the \spt{} coordinates adapted to 
the codimension-one space-like foliation 
 with the lapse \and{} shift \fnc{}s 
 $N =$$ \sqrt{g_{00}}$ \and{}  $N_i $$= g_{0i}=0$, \rsp{}ively,   
\and{} the induced metric $g_{ij}$ on the space-like leaves of the foliation.   
A detailed \trt{}ment of the \fnc{}al \Sch{} picture of QFT in flat \spt{} 
can be found in \cite{htf,jw}.

Instead, \prc{} \qnt{}ization of a scalar {}${}${} \fd{} $\phi(x)$ on a curved \spt{}  background 
 (cf.  \cite{ehr1,ehr2}) 
given \by{}  the metric tensor $g_{\mu\nu} (x)$  
 leads to the description  in \tmm{}s of a \clf{}-algebra-valued wave \fnc{} 
$\Psi(\phi, x^\mu)$ 
which satisfies 
the partial \drv{}ive \prc{}   \Sch{} \equn{} 
on the finite-dimensional bundle with the coordinates $(\phi,x^\mu)$: 
\beq  \label{crvns}
\ri\hbar \gma^\mu (x) 
 \nabla_\mu \Psi = 
 \left(- \frac{1}{2} \hbar^2\varkappa \frac{\partial^2}{\partial \phi^2 } 
+ \frac1\varkappa V(\phi)  \right)\Psi =: \frac1\varkappa\widehat{H}\Psi \,, 
\eeq
where $\gma$${}^\mu(x)$ are the Dirac matrices in curved space time which factorize the metric tensor $g$${}^{\mu\nu}(x)$: 
\beq \label{dmx}
\gma^\mu (x)\gma^\nu (x) + \gma^\nu (x) \gma^\mu (x) = 2g^{\mu\nu}(x), 
\eeq
$\nabla_\mu := \partial_\mu + \frac14$$\omega_\mu(x)$ is the covariant \drv{}ive 
acting on \clf{}-algebra-valued wave \fnc{}s, 
$\omega_\mu (x) {}= \omega_\mu{}_{AB}(x)\underline{\gma}{}^{AB}$ 
are the spin-\con{} matrices ${}$ (Fock--Ivanenko coefficients \cite{ivko,plk}), \and{} 
$\underline{\gma}{}_{A}$ are the constant 
tangent space Dirac matrices: $\underline{\gma}{}_{A} \underline{\gma}{}_{B} 
+ \underline{\gma}{}_{B} \underline{\gma}{}_{A}$$ = 2 \eta_{AB}$ \crsp{}ing to the 
Minkowski metric $\eta_{AB}$ with the ${}$ signature of our choice $+---$$...$.

The operator $\widehat{H}$ on the right-h\and{} {} side of (\ref{crvns}) is called the 
De Donder-Weyl (DW) \hlt{}ian {} operator \and{} it is constructed according to the procedure of 
\prc{} \qnt{}ization \cite{ehr1,ehr2}. 
It {} contains an ultraviolet parameter $\varkappa${} of the dimension of the inverse spatial volume 
which typically appears in the representations of \prc{} 
\qnt{}um operators already in flat {} \spt{} \cite{ehr1,qs96,bl,ldz}. 
 It is interesting to note that the \dwh{} operator $\widehat{H}$ 
coincides with the expression{}${}${} in flat \spt{} (cf. \cite{qs96,bl,ldz,ehr1}): 
the metric tensor components appear in the classical expression of the \dwh{} 
\fnc{} but they disappear in the \qnt{}um operator \crsp{}ing to it. The only manifestation of curved \spt{} is through the curved \spt{} Dirac matrices (\ref{dmx}) 
\and{} the spin-\con{} {}${}${}on the left-h\and{} side of {}Eq. (\ref{crvns}).

Obviously, the \prc{} description contrasts  with a familiar description of \qnt{}um \fd{}s derived from the canonical \qnt{}ization, in particular, with the description using the 
\fnc{}al \Sch{} picture {}${}${}outlined above, which explicitly distinguishes 
the role of space variables $\boldx$ as the labels of degrees of freedom \and{} the time 
variable $t$ along which the \qnt{}um evolution proceeds. 
In this paper, \myy{} will show {}${}${} that the description of  \qnt{}um scalar \fd{}s 
 on a static metric background in the \fnc{}al \Sch{} picture of QFT 
can be understood as a singular \lmt{}ing \cse{} of {}${}${}{}${}${}
 a 
\qnt{}um scalar theory in curved \spt{} 
derived from \prc{} \qnt{}ization.  
 This \lmt{}ing \cse{} {}${}${} \crsp{}s to the value of the ultraviolet parameter 
$\varkappa$ introduced \by{} \prc{} \qnt{}ization going to the regularized value of the 
Dirac delta \fnc{} at equal spatial points {}${}${}(which is the ultraviolet cutoff of 
the volume of the momentum space). 
This claim generalizes the {}${}${} similar result for  
\qnt{}um scalar \fd{}s \and{} \qnt{}um YM theory in flat \spt{} 
\cite{adv1,adv2,rmp18} (see also  \cite{my-pla,my-ym1} for an earlier \trt{}ment) 
to curved \spt{}. The insights \and{} notations 
from \cite{adv1,adv2,rmp18} will be closely follo\myy{}d {}${}${} 
in the present consideration. 
 
Let us recall that \prc{} \qnt{}ization \cite{qs96,bl,ldz,geom-q} 
is based on a generalization of the \hlt{}ian formalism to \fd{} theory known 
as the De Donder-Weyl (DW) \hlt{}ian theory \cite{dd,we,ru,ka}, which \trt{}s \spt{} 
variables on an equal footing. In this formulation,  
the Poisson brackets are defined on \dfrt{}ial forms representing the dynamical variables. 
The construction of the brackets uses the polysymplectic structure (whose integration over the initial data yields the standard symplectic 2-form 
on an infinite-dimensional phase space of a field theory) 
related to the Poincar\'e-Cartan form in the calculus of variations and it leads to the Poisson-Gerstenhaber algebra structure generalizing the usual Poisson algebra known in the canonical Hamiltonian formalism  \cite{geom-q,pg1,pg2}
 (for further generalizations see also \cite{joseph1,joseph2,my-loday,romer,my-dkp,my-dirac}). 
The existence of a \hlt{}-Jacobi theory \crsp{}ing to the \dwh{} formulation \cite{we,ru,ka} inevitably raises the question as to which formulation of the \qnt{}um theory of \fd{}s would reproduce the (partial \drv{}ive!) DW \hlt{}-Jacobi \equn{} in the classical \lmt{}. 
The 
precanonical \qnt{}ization aims at ${}$ clarifying this question. Instead of \qnt{}izing the 
full Poisson-Gerstenhaber algebra, it is based on \qnt{}ization of its small Heisenberg-like subalgebra, that leads to a hypercomplex generalization of the formalism of \qnt{}um theory where both operators \and{} wave \fnc{}s are \clf{}-algebra-valued, \and{} the \prc{} 
\Sch{} \equn{} includes the \spt{} Dirac operator which generalises the  time \drv{}ive in the st\and{}ard \Sch{} \equn{}\cite{qs96,bl,ldz,geom-q}. 
 One of the features of this formulation of \qnt{}ized  \fd{}s is that it allows us to reproduce the classical \fd{} \equn{}s as the \equn{}s of expectation values of operators defined \by{} \and{} evolving according to the \equn{}s of \prc{} \qnt{}ization \cite{ehr1,ehr2}. Moreover, \by{} \trt{}ing the \spt{} variables on an equal footing \and{} leading to a formulation on a finite-dimensional space of \fd{} \and{} \spt{} variables the \prc{} \qnt{}ization approach provides a natural \and{} promising framework for  \qnt{}ization of gravity \cite{ijtp2001,qg} \and{} even for the consideration  of the mass gap problem in \qnt{}um gauge theory \cite{my-massgap}. 
On the classical level, the \dwh{} formulation \and{} the underlying polysymplectic structure have been used recently in order to construct  highly effective numerical 
integrators for PDEs \and{} general relativity \cite{new38,new39}. 

In this paper, in Sec. 2 \myy{} assume that the \Sch{} wave \fnc{}al $\Psbld$ is a \fnc{}al of \prc{} 
wave \fnc{} $\Psi$ \rsd{} to a \fd{} configuration $\phi(\boldx)$ at time $t$. This \imls{} 
a global space+time decomposition in the \prc{} \Sch{} \equn{}  (\ref{crvns}). The time evolution 
of the wave \fnc{}al $\Psbld$ is then controlled \by{} the time evolution of the \prc{} wave \fnc{} $\Psi$ \rsd{} to the  subspace $\Si$ representing the above mentioned \fd{} configuration: $\Psi_\Si$. We will confine ourselves to the \cse{} of static \spt{}s \and{} present a restriction of the \prc{} \Sch{} \equn{} (\ref{crvns}) to $\Si$, 
which controls the time evolution of the \prc{} wave \fnc{} \rsd{} to $\Si$.   
 Then \myy{} will use it to write an \equn{} for the time evolution of an arbitrary wave \fnc{}al of $\Psi_\Si$. By{} comparing the \tmm{}s in this \equn{} with the expressions of variational \drv{}ives 
of the composite \fnc{}al $\Psbld([\Psi_\Si(\phi(\boldx), t)])$ with \rsp{} to $\phi(\boldx)$ 
\myy{} demonstrate how the \tmm{}s in the \fnc{}al \Sch{} \equn{} are emerging from the \tmm{}s in the \prc{} \Sch{} \equn{} \wnn{} the parameter $\varkappa$ in the \ltr{} is replaced \by{} a singular scalar combination of Dirac $\dlt$-\fnc{} at equal spatial points, $\gma_0$ matrix \and{} the density $\sqrt{-g}$. In this  \lmt{}ing \cse{}, \myy{} will be able to \ota{} an expression of the Sch\"odinger wave \fnc{}al in \tmm{}s of the \prc{} wave \fnc{} \and{} to argue that the contributions from the additional \tmm{}s in the \equn{} for $\Psi_\Si$, which have no counterparts in the \fnc{}al \drv{}ive \Sch{} \equn{} (\ref{fuse}), are \vsh{}ing. In this way \myy{} derive (\ref{fuse}) from (\ref{crvns}). Our conclusions \and{} some problems for further work are presented in Sec. 3.

\section{Schr\"{o}dinger wave functional from precanonical wave function }

In the \cse{} of interacting scalar \fd{}s in flat \spt{},  
the \con{} \bewn{} the \fnc{}al \Sch{} 
representation \and{} \prc{} \qnt{}ization was established in 
\cite{adv1,adv2}. It was further extended to \qnt{}um YM theory in flat \spt{} in \cite{rmp18}.  These papers improve 
our earlier considerations in \cite{my-pla} \and{} \cite{my-ym1}.  
The  conclusion is 
that the st\and{}ard QFT in \fnc{}al \Sch{} representation can be derived from 
the \prc{} \qnt{}ization in the \lmt{}ing \cse{} of the infinite parameter $\varkappa$ or, 
more precisely,  \wnn{} the combination $\underline{\gma}{}_0\varkappa$ is replaced \by{} $\dlt(\mathbf{0})$, 
a regularized value of Dirac delta \fnc{} $\dlt(\boldx-\boldx')$ at 
coinciding spatial points, i.e. the cutoff of the momentum space volume introduced \by{} regularization. 

 Here \myy{} explore a similar \con{} \bewn{} the \fnc{}al 
Schr\"o\-dinger representation of a \qnt{}um scalar \fd{} theory in curved \spt{} 
\and{} \prc{} \qnt{}ization of the same system.  
Specifically, \myy{} restrict ourselves to static \spt{}s. In this \cse{}, 
$\partial_t g^{\mu\nu}=0$, all curved \spt{} Dirac matrices 
$\gma^\mu$ are $\boldx$-dependent, \and{} the consistency 
with the choice of the adapted coordinate system used in (\ref{fuse}) 
\imls{} for the Christoffel symbols \and{} the spin-\con{} coefficients 
$\Gamma_{0\nu}^\mu = 0 = \Gamma_{\mu\nu}^0 $ \and{} $\omega_0^{IJ} = 0 = \omega_i^{0J}$, \rsp{}ively.  
An extension of the consideration below to more general \spt{}s will be 
presented elsewhere \cite{new40}.

Intuitively, the \rln{} \bewn{} the {}${}${}\Sch{} wave \fnc{}al \and{} the \prc{} wave \fnc{} is suggested \by{} the probabilistic interpretation. While the former has the meaning of the probability amplitude of finding a \fd{} configuration $\phi(\boldx)$ at some moment of time $t$, {}${}${}
the \ltr{} can be interpreted  as the probability amplitude of observing the \fd{} value $\phi$ at the \spt{} point $x$. This allows us to expect that the time-dependent  complex \fnc{}al probability amplitude $\Psbld([\phi(\boldx)], t)$ is a composition of \spt{} dependent \clf{}-valued probability {}${}${}amplitudes given \by{} the \prc{} wave \fnc{}  $\Psi(\phi,x)$.

A  \rln{} \bewn{} the \fnc{}al \Sch{} picture \and{} 
the \prc{} descripton {}${}${} \imls{} that 
 the  \Sch{} wave \fnc{}al $\Psbld$$ ([\phi(\boldx)],t)$ 
 can be expressed as a \fnc{}al of \prc{} wave \fnc{}s $\Psi$$ (\phi, x)$
\rsd{} to a specific \fd{} configuration $\Si$: $\phi=\phi(\boldx)$ at time $t$:  
\beq \label{psib}
\Psbld([\phi(\boldx)],t) = \Psbld ([\Psi_\Si (\boldx,t), \phi (\boldx)]) \,,  
\eeq
where \myy{} have \dnt{}d the restriction of \prc{} wave \fnc{} $\Psi(\phi,x)$  to $\Si$ as 
$\Psi_\Si (\boldx,t) := \Psi (\phi=\phi(\boldx), \boldx, t)$. 
Thus the time dependence of the wave \fnc{}al $\Psbld$ is totally controlled \by{} the 
time dependence of \prc{} wave \fnc{} \rsd{} to $\Si$ \and{} the 
time \drv{}ive of $\Psbld$ is \ota{}ed \by{} the chain rule \dfrt{}iation 
\beq \label{dtpsi0}
 \ri\partial_t \Psbld = 
 {\Tr} \int\! d\boldx\, 
 \left \{ 
 \frac{\dlt \Psbld }{\dlt\Psi^T_\Si(\boldx, t)} 
\ri\partial_t \Psi_\Si (\boldx, t)  
\right \} \,, 
 \eeq
where $\Psi^T$ is the transpose matrix of $\Psi$. In what follows, for brevity, \myy{} will 
\dnt{} $\Psi_\Si (\boldx, t)$  simply as $\Psi_\Si (\boldx)$ or even $\Psi_\Si $ \wnn{}  appropriate.

\medskip

The time \drv{}ive $\ri\partial_t \Psi_\Si (\boldx, t)$ is de\tmm{}ined \by{} the restriction of 
\prc{} Schr\"o\-dinger \equn{} (\ref{crvns}) 
to  $\Si$, which takes the form 
\beqa \label{psisigmaeqn}
\ri\partial_t \Psi_\Si = -\ri\gma_0\gma^i \left ( \frac{d}{dx^i} 
- \partial_{i} \phi (\boldx) \frac{\partial}{\partial \phi } \right ) \Psi_\Si 
- \frac{\ri}{4} \gma_0 \gma^{i} [\omega_i, \Psi_\Si ]  
+ \gma_0 \frac1\varkappa \widehat{H}_\Si \Psi_\Si~, 
\eeqa 
where  $\frac{d}{dx^i} $ is the total \drv{}ive along $\Si$: 
\beq \label{ttl}
\frac{d}{dx^i}:= \partial_i + \partial_i \phi(\boldx)  \frac{\partial}{\partial \phi} 
+ \partial_i \phi_{,k}(\boldx) \frac{\partial}{\partial \phi_{,k} } +... , 
\eeq
$\phi_{,k}$ \dnt{} the fiber coordinates of the 
\fst{}-jet bundle of the bundle of \fd{} varibles $\phi$ over \spt{} 
(cf. \cite{saunders,olver}) 
\and{} $\widehat{H}_\Si$ is the restriction  of the 
\dwh{} operator to $\Si$: 
\beq \label{dwhop}
\frac1\varkappa\widehat{H}_\Si =  \frac1\varkappa\widehat{H} = 
-  \frac{\varkappa}{2}  \frac{\partial}{\partial \phi^2 } 
+ \frac1\varkappa V(\phi) .
\eeq 
In (\ref{psisigmaeqn})  \myy{} already explicitly specify the action of the spin-\con{} 
on \clf{}-valued \fnc{}s via a commutator, though the necessity of it will be revealed 
only later in Eq.~(\ref{first}). 
The \fst{} two \tmm{}s in (\ref{psisigmaeqn}) can be 
 rewritten in \tmm{}s of the  
total covariant \drv{}ive $\nabla^{\mathrm{tot}}$ 
 of $\Psi$ \rsd{} to $\Si$, 
\beq \label{27}
\ri\partial_t \Psi_\Si = -\ri\gma_0\gma^i \nabla^{{\mathrm{tot}}}_i \Psi_\Si  
 + \ri\gma_0\gma^i\partial_{i} \phi (\boldx) \partial_\phi\Psi_\Si 
 +  \frac1\varkappa \gma_0\widehat{H}_\Si \Psi_\Si~, 
\eeq
where the superscript ${}^{{\mathrm{tot}}}$ has two meanings: the \fst{} meaning 
is that the \drv{}ive is total in the sense of being 
taken of a composite \fnc{} $\Psi(\phi(\boldx),\boldx)$, \and{} the second meaning is that the covariant 
\drv{}ive of a \clf{}-valued tensor \fnc{} $T^{\mu_1\mu_2...}_{\nu_1\nu_2...}$ 
is total in the sense that it includes both the spin-\con{} 
 $\omega_\mu{}^{IJ}$ \and{} the Christoffel symbols $\Gamma^\alpha_{\beta\gma}$: 
\beq \label{cov}
\nabla^{{\mathrm{tot}}}_\alpha T^{\mu_1\mu_2...}_{\nu_1\nu_2...} := 
\frac{d}{d x^\alpha} T^{\mu_1\mu_2...}_{\nu_1\nu_2...} 
+ \frac14 [\omega_\alpha, T^{\mu_1\mu_2...}_{\nu_1\nu_2...}] 
+ [\Gamma , T]^{\mu_1\mu_2...}_{\alpha\nu_1\nu_2...} , 
\eeq  
where $[\Gamma , T]^{\mu_1\mu_2...}_{\alpha\nu_1\nu_2...}$ is just a short-h\and{} notation for 
$\Gamma^{\mu_1}_{\alpha \beta} T^{\beta \mu_2...}_{\nu_1\nu_2...} 
+ \Gamma^{\mu_2}_{\alpha \beta} T^{\mu_1\beta ...}_{\nu_1\nu_2...}  +.... 
- \Gamma^{\beta}_{\alpha \nu_1} T^{\mu_1 \mu_2 ...}_{ \beta\nu_2...}  
 - \Gamma^{\beta}_{\alpha \nu_2} T^{\mu_1 \mu_2 ...}_{ \nu_1 \beta...}
- ... $ 
\and{} the commutator with the spin-\con{} matrix in the second \tmm{} ensures 
that the covariant \drv{}ive satisfies the Leibniz rule \wnn{} acting on the \clf{} (matrix) product of two \clf{}-valued tensor \qnt{}ities. This property will be very important for our integrations  \by{} parts below.  
The last \tmm{} in (\ref{cov}) \vsh{}es \wnn{} acting on the scalar \clf{}-valued \fnc{} 
$\Psi$. Ho\myy{}ver, it appears in the 
 condition of the covariant constancy of Dirac matrices $\gma^\mu$ \and{} their antisymmetric products $\gma^{\mu_1 \mu_2...}$, which in \tmm{}s  of $\nabla^{{\mathrm{tot}}}_\alpha$ reads 
\beq  \label{mcomp}
\nabla^{{\mathrm{tot}}}_\alpha \gma^{\mu_1 \mu_2...} = 0. 
\eeq 
Obviously, \wnn{} acting on $x$-dependent $\gma$-matrices, only the \fst{} partial \drv{}ive \tmm{} 
in (\ref{ttl}) is non-\vsh{}ing. 

\medskip 
 
 Now, from (\ref{dtpsi0}), (\ref{psisigmaeqn}) \and{} (\ref{dwhop}) the time evolution of the wave \fnc{}al of the \qnt{}um scalar \fd{} in static \spt{} is given \by{} 
\begin{align} \label{dtbpsi1}
\begin{split}
i\partial_t \Psbld =
\int \!\rd\boldx\ 
{\sf Tr} \Big\{
 \BPhi^{}({\boldx}) 
&\Big[ 
\underbrace{-\ri\gma_0\gma^i \frac{d}{dx^i} \Psi_\Si  (\boldx)}_{I} 
 + \underbrace{\ri \gma_0\gma^i\partial_i \phi(\boldx){\partial_\phi} \Psi_\Si  (\boldx)
 \vphantom{\frac{A}{B}} }_{II}
   \\
 \underbrace{\strut - \frac{\ri}{4} \gma_0 \gma^{i} [\omega_i{}, \Psi_\Si (\boldx)]
 \vphantom{\frac{A}{B}} }_{III}  &
\underbrace{- \frac{\varkappa}{2} \gma_0 \partial_{\phi\phi}\Psi_\Si  (\boldx)}_{IV} 
+
  \underbrace{\frac{1}{\varkappa}\gma_0 V(\phi(\boldx)) \Psi_\Si  (\boldx)}_{V}
\Big ] \Big \} , 
\end{split}
\end{align} 
where the notation 
\beq \label{defphi}
\BPhi^{}({\boldx})
:= \frac{\dlt \Psbld}{\dlt\Psi^T_\Si (\boldx)} 
\eeq 
is introduced. 
We would like to see how this \equn{} can reproduce the \fnc{}al \drv{}ive 
\Sch{} \equn{} (\ref{fuse}).

 In order to compare (\ref{dtbpsi1}) with (\ref{fuse}), let us calculate 
  the \fst{} \and{} the second
total \fnc{}al \drv{}ives of  $\Psbld$ in  (\ref{psib}) with \rsp{} to $\phi(\boldx)$: 
\begin{align} 
\label{delta-bpsi}
 \frac{\dlt \Psbld }{\dlt \phi(\boldx)}
&=
{\sf Tr}\left \{
 \BPhi^{}({\boldx})
\partial_\phi \Psi_\Si (\boldx)
\right \}
+  \frac{\dltb \Psbld }{\dltb \phi(\boldx)^{{}^{}}}, \\[1ex]
\begin{split}
\label{delta-bpsi2} 
\frac{\dlt^2 \Psbld }{\dlt \phi(\boldx)^2}
&=
{\sf Tr}\left \{
 \BPhi^{}({\boldx}) 
~\dlt(\mbox{\bf 0})\partial_{\phi\phi} \Psi_\Si (\boldx) 
+ 2 \frac{ \dltb \BPhi(\boldx)}{\dltb \phi(\boldx)}   ~\partial_\phi \Psi_\Si (\boldx) \right \}
+  \frac{\dltb^2 \Psbld }{\dltb \phi(\boldx)^2} 
  \\[1ex]
 &\quad+{\sf Tr} \, {\sf Tr} \left \{
 \frac{\dlt^2 \Psbld}{\dlt \Psi^T_\Si(\boldx)\otimes\dlt\Psi^T_\Si(\boldx)}
~\partial_\phi \Psi_\Si (\boldx)
\otimes  \partial_\phi \Psi_\Si  (\boldx)
\right \} .
\end{split}
\end{align} 
where the double trace notation in the last line refers to the fact that the trace is taken of each 
of the matrices in the direct product. 
Here \and{} in what follows,  $\dltb$ \dnt{}s
 the 
partial \fnc{}al \drv{}ive
with \rsp{} to $\phi(\boldx)$,  
 \and{} 
$\dlt(\mbox{\bf 0})$ is the  $(n-1)$-dimensional delta-\fnc{} at $\boldx=0$ 
 which results from the variational \dfrt{}iation 
of the \fnc{} $\Psi_\Si (\boldx)$ with \rsp{} to itself at the same spatial point. 
Henceforth, \wnn{} writing $\dlt(\mbox{\bf 0})$, 
\myy{} imply that a 
proper regularization like a point splitting  or a lattice one has been applied in order 
to make sense of this singular expression. This is the regularization which is usually implied \wnn{} the second variational \drv{}ive is used in the \fnc{}al \Sch{} \equn{} in QFT.

\medskip

Let us start from the observation that the potential energy \tmm{} in the canonical \fnc{}al 
\drv{}ive \Sch{} \equn{} for the \qnt{}um scalar \fd{} (\ref{fuse}) should be \ota{}ed from the potential energy \tmm{} $V$ 
 in (\ref{dtbpsi1}),  i.e.
\beq
 \int \!\rd\boldx\ 
 {\sf Tr}
\left \{
\BPhi(\boldx)
~\frac{1}{\varkappa}
\gma_0
V(\phi(\boldx))\Psi_\Si (\boldx)) \right \}
\mapsto \int \!\rd\boldx \sqrt{-g}\ V(\phi(\boldx))~\Psbld , 
\eeq
where a more precise meaning of the symbol $\mapsto$  will be clarified below. 
  To accomplish that, the following \rln{} should be fulfilled at  any spatial point $\boldx$: 
\beq \label{v-term}
{\sf Tr} \left \{    
\BPhi(\boldx) 
 ~ \frac{1}{\varkappa} \gma_0 
 \Psi_\Si (\boldx) \right \}
  \mapsto \sqrt{-g}\ \Psbld .
\eeq
 By
 \fnc{}ally \dfrt{}iating both sides of Eq. (\ref{v-term}) with \rsp{} to
$\Psi^T_\Si (\boldx)$, \myy{} \ota{} 
\beq \label{kade}
{\sf Tr} \left \{
\frac{\dlt^2 \Psbld}{\dlt\Psi^T_\Si (\boldx)
\otimes \dlt\Psi_\Si (\boldx)}
 \frac{1}{\varkappa} \gma_0 \Psi_\Si (\boldx)  \right \}
+ 
 \BPhi(\boldx) 
 \frac{1}{\varkappa}\gma_0 \dlt(\mbox{\bf 0})
 \mapsto
 \sqrt{-g}\ \BPhi(\boldx)  , 
\eeq
where
$\dlt(\mbox{\bf 0}) :={\dlt \Psi_\Si(\boldx)}/{\dlt \Psi_\Si(\boldx)}.$
 This type of \rln{} is possible only if 
 the second variational \drv{}ive of $\Psbld$ with \rsp{} to $\Psi_\Si(\boldx)$ 
 \vsh{}es:  
 \beq \label{deltapsipsi}
\frac{\dlt^2 \Psbld}{\dlt\Psi_\Si (\boldx)
 \otimes  \dlt\Psi_\Si (\boldx) } =0  , 
\eeq
 \and{} 
\beq \label{aaa}
\frac{1}{\varkappa} \gma_0(\boldx) \dlt(\mbox{\bf 0}) -  \sqrt{-g} (\boldx) 
\mapsto 0  
\eeq
(here \myy{} explicitly recall that both $\gma_0$ \and{} $\sqrt{-g}$ depend on $\boldx$-s).  
The \ltr{} can be understood as the condition 
\beq  \label{lim}
 \gma^0  \sqrt{-g}   \varkappa \mapsto \dlt(\mbox{\bf 0}) . 
\eeq 
By noticing that $\sqrt{-g} = \sqrt{- g_{00} h}$, where $h:=\det ||g_{ij}||$ 
is the de\tmm{}inant of the spatial part of the metric tensor, \and{} $\gma^0 \sqrt{g_{00}} = 
\underline{\gma}{}_0$ is the time-like component of the tangent Minkowski space Dirac matrices, \myy{} can rewrite (\ref{lim}) in the form 
\beq \label{limh}
\underline{\gma}{}_0\varkappa \mapsto \dlt(\mathbf{0})/\sqrt{-h} = \dlt^\mathrm{inv}(\mathbf{0}) ,
\eeq 
where in the last equality the invariant $(n-1)$-dimensional delta \fnc{} appears 
which is defined \by{} the property $\int\! d\boldx \sqrt{-h}(\boldx)\dlt^\mathrm{inv}(\mathbf{x}) =1$. 
We, therefore, see that the curved \spt{} generalization of the \lmt{}ing map 
$\underline{\gma}{}_0\varkappa
\mapsto\dlt(\mbox{\bf 0})$ found earlier in flat \spt{} \cite{adv1,adv2} 
just replaces the $(n-1)$-dimensional delta \fnc{} \by{} the invariant one, at least in static \spt{}s.

\medskip

 Similarly,  in the \lmt{}ing \cse{} ({\ref{lim}}),
the \tmm{} 
$IV$ in  (\ref{dtbpsi1}) reproduces the \fst{} \tmm{} on the right-h\and{} side of (\ref{delta-bpsi2}): 
\beq
IV: \quad -\frac{\varkappa}{2}\gma_0 \partial_{\phi\phi}\Psi_\Si 
\mapsto   - \frac{1}{ \sqrt{-g}} g_{00} \dlt({\mathbf{0}})\partial_{\phi\phi}\Psi_\Si .
\eeq
Then, \by{} comparing with  (\ref{delta-bpsi2})  \myy{} conclude that the \tmm{} $IV$ in  (\ref{dtbpsi1}) 
 leads to the following expression in variational \drv{}ives of $\Psbld$: 
\begin{align} \label{e218}
\begin{split} 
IV\!: \; 
{\sf Tr} \Big\{
\frac12 
& \BPhi (\boldx)\varkappa\gma_0 \partial_{\phi\phi}\Psi_\Si  (\boldx) \Big\} 
\\ 
&\mapsto 
\frac12 \frac{g_{00}}{\sqrt{-g}}  
\left( \frac{\dlt^2 \Psbld }{\dlt \phi(\boldx)^2} 
 - 2~{\sf Tr} \left \{
\frac{\dltb \BPhi(\boldx)}{\dltb \phi(\boldx)}
~\partial_\phi \Psi_\Si (\boldx) \right \} 
 -  \frac{\dltb^2 \Psbld }{\dltb \phi(\boldx)^2} \right) . 
\end{split}
\end{align}
While the \fst{} \tmm{} on the right-h\and{} side of (\ref{e218}) correctly reproduces the  \fst{} \tmm{} in the \fnc{}al \drv{}ive \Sch{} \equn{} (\ref{fuse}), two other \tmm{}s need further investigation, which follows.

\medskip

Since \myy{} expect  Eq. (\ref{dtbpsi1}) to lead to a description solely in \tmm{}s of the wave \fnc{}al $\Psbld$,  as in the \fnc{}al \Sch{} \equn{} (\ref{fuse}), 
the \tmm{} $II$ in (\ref{dtbpsi1}) with $\partial_\phi \Psi_\Si$ should be cancelled \by{} some other 
\tmm{} with $\partial_\phi \Psi_\Si(\boldx)$. Now \myy{} can do so  
with the help of the second \tmm{} in ({\ref{e218}}). Thus the requirement of mutual 
cancellation of the \tmm{}s with $\partial_\phi \Psi_\Si(\boldx)$ leads to the \rln{} 
\beq \label{dypsi} 
II + \mathrm{part\,of \,} IV:\quad 
 \BPhi^{}({\boldx})
 \ri\gma_0 \gma^i\partial_i\phi(\boldx)
 +  \frac{g_{00}}{\sqrt{-g}} 
\frac{\dltb \BPhi (\boldx)}{\dltb \phi(\boldx)} \mapsto 0.
\eeq
By taking the variational \drv{}ive of (\ref{dypsi})  with \rsp{} to  $\phi(\boldx')$ 
\myy{} can easily see that (\ref{dypsi})   with $\mapsto$ replaced \by{} the equality sign 
is not an integrable \equn{} in variational \drv{}ives. Nevertheless, \by{} taking into account that it should be valid only under the 
\lmt{}ing map (\ref{lim}), \myy{} can write the solution for $\BPhi(\boldx)$ in the form 
\beq \label{bphi}
\BPhi^{}(\boldx) =  \BXi([\Psi_\Si];\check{\boldx})
e^{-\ri \phi(\boldx)\gma^i\partial_i\phi(\boldx)/\varkappa},
\eeq
where $\BXi([\Psi_\Si];\check{\boldx})$ \dnt{}s
 a \fnc{}al of $\Psi_\Si (\boldx')$ at $\boldx'\neq \boldx$ (i.e. on a punctured space with the removed point $\boldx$), 
i.e. $\frac{\dltb \BXi([\Psi_\Si];\check{\boldx})}{\dltb \phi(\boldx)}\equiv 0$, 
which plays the role of the integration constant here. 
 Indeed, \by{} \dfrt{}iating (\ref{bphi}) with \rsp{} to $\phi(\boldx)$, 
 replacing $\varkappa$ according to the \lmt{}ing map (\ref{lim}), \and{} taking into account that 
 $\partial_i\dlt(\mathbf{0})=0$ \and{} 
 $\gma_0(\boldx)\gma_0(\boldx) = g_{00}(\boldx)$,  
 \myy{} conclude that (\ref{dypsi}) is fulfilled \by{} (\ref{bphi}) under the condition (\ref{lim}).  
 Moreover, 
  (\ref{bphi}) \by{} construction fulfills 
\beq 
\frac{\dlt \BPhi (\boldx )}{\dlt \Psi^T_\Si (\boldx )} 
= \frac{\dlt^2 \Psbld}{\dlt\Psi^T_\Si (\boldx)
 \otimes  \dlt\Psi^T_\Si (\boldx) } \equiv 0 , 
\eeq
which is consistent with (\ref{deltapsipsi}) as it should. 
Let us emphasize that the cancellation of the \tmm{}s with 
$\partial_\phi \Psi_\Si(\boldx)$ can be only achieved in the \lmt{}ing \cse{} (\ref{lim}).

\medskip 

Now, based on (\ref{defphi}), (\ref{v-term}) \and{} (\ref{bphi}) \myy{} can write the wave \fnc{}al $\Psbld$  
 in the following form valid at any point  $\boldx$: 
\beq \label{bpsi3}
\Psbld \sim  
 {\sf Tr}
\left \{\BXi([\Psi_\Si];\check{\boldx})~
e^{-\ri \phi(\boldx)\gma^i\partial_i\phi(\boldx)/\varkappa}~
 \frac{\gma_0}{\sqrt{-g} \varkappa}
 \Psi_\Si (\boldx) \right \}_{\mbox{\Large $\rvert$} \scriptstyle
 \varkappa \stackrel{\mbox{\tiny $$}}{\mbox{\scriptsize $\longmapsto$}} 
 \gma_0\dlt(\mathbf{0})/ \sqrt{-g}  },
\eeq  
where $\sim$ 
\dnt{}s the equality up to a normalization factor, which will also include $\varkappa$ \and{} $\sqrt{-g}$,  
\and{} the notation $\{ ... \}_{\mbox{\large $\rvert$} \scriptstyle
 \varkappa \stackrel{\mbox{\tiny $$}}{\mbox{\scriptsize $\longmapsto$}} 
 \gma_0\dlt(\mathbf{0}) / \sqrt{-g}}$
 means that every appearance of $\varkappa$ in the expression inside braces 
 is replaced \by{}  
 $\gma_0\dlt(\mathbf{0}) / \sqrt{-g}$ with a regularized value of $\dlt(\mathbf{0})$ 
 in accord with the \lmt{}ing map (\ref{lim}).  

Using this expression for the wave \fnc{}al $\Psbld$ \myy{} can now evaluate the last \tmm{} 
in (\ref{e218}) in the \lmt{} (\ref{lim}): 
\beq 
\mathrm{part\,of \,} IV:\quad 
\frac12 \frac{g_{00}}{\sqrt{-g}} \frac{\dltb^2 \Psbld }{\dltb \phi(\boldx)^2}
\mapsto
 -  \frac{1}{2}\sqrt{-g} g^{ij}\partial_i \phi(\boldx) \partial_j \phi(\boldx) \Psbld .  
\eeq
We see that it correctly reproduces the second \tmm{} in the \fnc{}al \drv{}ive Schr\"o\-dinger \equn{} (\ref{fuse}) which is responsible for the inherent non-ultralocality 
(adapting  Klauder's \tmm{}inology \cite{klauder}) of  \qnt{}um relativistic scalar \fd{} theory.  

Thus, in the \lmt{}ing \cse{} (\ref{lim}), 
\myy{} have successfully derived all \tmm{}s in (\ref{fuse}) from our \prc{} 
\Sch{} \equn{} \rsd{} to $\Si$, Eq. (\ref{psisigmaeqn}). 
Ho\myy{}ver, there are still two  \tmm{}s $I$ \and{} $III$ left in (\ref{dtbpsi1}) 
 which have not played any role yet: 
 \beq \label{227}
 I+III:\quad -\ri \int \!\rd\boldx\ {\sf Tr} \Big\{\BPhi(\boldx) \gma_0\gma^i \nabla^{{\mathrm{tot}}}_i\Psi_\Si 
  \Big\} .
 \eeq
In our previous discussions in flat \spt{} \cite{adv1,adv2,rmp18}
it has been always easy to show in the end of the calculation that the \crsp{}ing 
\tmm{} with the total \drv{}ive 
$d/ d x^i \Psi_\Si (\boldx)$ is \vsh{}ing (under the boundary condition that 
$\Psi_\Si (\boldx)$ is \vsh{}ing at the spatial infinity). 
Let us see if this property extends to the \cse{} of static curved \spt{}s which \myy{} consider here.  
 
 By integration
\by{} parts using the covariant Stokes theorem \and{} the Leibniz property of the covariant \drv{}ive 
$\nabla^{{\mathrm{tot}}}$  acting on \clf{}-valued \fnc{}s, (\ref{227}) can be transformed as follows: 
\begin{align} \label{first} 
I+III:\quad 
&-\ri \int \!\rd\boldx\ \sqrt{-h}\
\left( {\sf Tr} \Big\{ \frac{1}{\sqrt{-h}}
 \BPhi(\boldx) \gma_0\gma^i \nabla^{{\mathrm{tot}}}_i\Psi_\Si \Big\}  
 \right) 
\nn \\ = & 
-\ri \int \!\rd\boldx\ \sqrt{-h}  {\sf Tr} 
\Big\{ \nabla^{{\mathrm{tot}}}_i \Big(\frac{1}{\sqrt{-h}}
 \BPhi(\boldx) \gma_0\gma^i  \Psi_\Si \Big)\Big\} 
 \nn \\ 
  &+ \ri \int \!\rd\boldx\ \left( \sqrt{-h} {\sf Tr} 
\Big\{\nabla^{{\mathrm{tot}}}_i \Big(\frac{1}{\sqrt{-h}}
 \BPhi(\boldx) \gma_0\gma^i\Big) \Psi_\Si \Big\} 
 \right)
\nn \\
 \begin{split}
 = & -\ri \oint_{\partial \Si} \!\rd\boldx_i 
  {\sf Tr} \Big\{ \BPhi \gma_0\gma^i \Psi_\Si \Big\} 
+  \ri \int \!\rd\boldx\  {\sf Tr} 
\Big\{ \BPhi \big(\nabla^{{\mathrm{tot}}}_i ( \gma_0\gma^i) \big) \Psi_\Si \Big\}
 \\ +& \ri \int \!\rd\boldx\ 
 \left( \frac{-\nabla_i \sqrt{-h}}{\sqrt{-h}}
{\sf Tr}\big\{ \BPhi  \gma_0\gma^i\Psi_\Si \big\} 
+  {\sf Tr} 
\big\{ \big(\nabla^{{\mathrm{tot}}}_i 
 \BPhi(\boldx) \big) \gma_0\gma^i \Psi_\Si \big\} \right.
 \bigg) . 
  \end{split}
  \end{align} 
Here, 
\begin{itemize}
\item[(i)]
the \fst{} boundary \tmm{} on the right-h\and{} side of (\ref{first}) follows from the covariant Stokes theorem;\ 
 $\rd \boldx_i = d^{n-2} \boldx|_{\partial\Si} n_i (\boldx) $ 
is the measure of $(n-2)$-dimensional integration  
on the boundary $\partial\Si$ with the normal vector $n_i (\boldx)$ tangent to $\Si$. 
Under the assumption  that $\Psi$ \vsh{}es on the boundary $\partial \Si$ 
the boundary \tmm{} \vsh{}es too. 
\item[(ii)] 
 Next three \tmm{}s on the right-h\and{} side of (\ref{first}) follow from the 
Leibniz rule with \rsp{} to the \clf{} product fulfilled \by{} the total 
covariant \drv{}ive $\nabla^{{\mathrm{tot}}}_i$ acting on tensor \clf{}-algebra-valued \fnc{}s. 
This is where the fact that the 
spin-\con{} matrix $\omega_i$ acts on $\Psi$ \by{} means of the commutator product 
$[\omega_i, \Psi]$ is essential.  
\item[(iii)] 
The second \tmm{} with 
$\nabla^{{\mathrm{tot}}}_i ( \gma_0\gma^i) 
$ \vsh{}es 
due to the covariant constancy of Dirac matrices (\ref{mcomp}). 
\item[(iv)]
The third \tmm{} \vsh{}es because the covariant \drv{}ive of the density 
$\sqrt{-h}$ \vsh{}es: $\nabla_i \sqrt{-h} = 0$, as a consequence of the covariant 
constancy of $g_{\mu\nu}$. 

\item[(v)] 
In the fourth \tmm{} on the right-h\and{} side of (\ref{first}), 
using the explicit formula for $\BPhi (\boldx) $ in (\ref{bphi}) \and{} the Leibniz rule for $\nabla^{{\mathrm{tot}}}_i$, \myy{} \ota{} 
\beq \label{229}
\nabla^{{\mathrm{tot}}}_i \BPhi (\boldx) = \frac{-\ri}{\varkappa}\BPhi (\boldx) \big(\partial_i \phi\gma^l\partial_l \phi 
+ \phi \gma^l \partial_{il} \phi + \phi (\nabla^{{\mathrm{tot}}}_i \gma^l) \partial_l \phi \big).
\eeq
The last \tmm{} in (\ref{229}) \vsh{}es due to the covariant constancy of Dirac matrices (\ref{mcomp}). By substituting (\ref{229}) into the last \tmm{} in (\ref{first}), 
integrating \by{} parts (assuming the \fd{} configurations $\phi(\boldx)$ \vsh{}  at the spatial infinity) \and{} using the covariant Stokes theorem again, \myy{} \ota{} 
\begin{align} 
\begin{split} \label{230}
 \int \!\rd\boldx\ {\sf Tr} \Big\{  & ||\Psbld|| \sqrt{-g}
\big( g^{il}\partial_i \phi \partial_l\phi + \phi g^{il}\partial_{il}\phi \big)  \Big\} 
\\ &= 
- \Psbld \int \!\rd\boldx\ \sqrt{-h} \nabla^{}_i\big(\sqrt{g_{00}}g^{il}\big)\frac12 \partial_l \phi^2
=0,  
\end{split}
\end{align}
where $||\Psbld||:=  \BPhi(\boldx)  \frac{1}{\varkappa \sqrt{-g}} \gma_0 \Psi_\Si (\boldx)$ 
such that  ${\sf Tr}  ||\Psbld|| = \Psbld$ (c.f. (\ref{v-term})) is independent of $\boldx$.    
Again, the right-h\and{} side of (\ref{230}) \vsh{}es because the covariant \drv{}ive of the metric tensor \vsh{}es. 
\end{itemize}

Thus \myy{} have shown that all four \tmm{}s on the right-h\and{} side of (\ref{first}) \vsh{} \and{}, 
therefore, the \tmm{}s $I$ \and{} $III$ in (\ref{dtbpsi1}) in the \lmt{}ing \cse{} (\ref{lim}) 
yield a \vsh{}ing contribution in the \equn{} for the \fnc{}al $\Psbld$.  

Note that this result is a consequence of the properties of the pseudo-Riemannian geometry of the \spt{} background,
the assumed boundary conditions that the values of $\Psi_\Si(\boldx)$ \and{} $\phi(\boldx)$ at $\boldx \rightarrow \infty$ are \vsh{}ing, \and{} the particular form of the \fnc{}al $\BPhi(\boldx)$  which was established earlier in (\ref{bphi}).  The covariant Stokes theorem \and{} the 
Leibniz property of the covariant \drv{}ive with \rsp{} to the \clf{} product 
have been instrumental. 
The \ltr{} property is guaranteed only \wnn{} the 
spin-\con{} matrix acts on the \clf{}-algebra-valued wave \fnc{}s \by{} the 
commutator product. 

\medskip

Finally, \myy{} can specify the \fnc{}al $\BXi([\Psi_\Si(\bx)], \check{\boldx})$ in (\ref{bpsi3}) \by{} combining all the above observations together \and{} noticing that the 
formula Eq.~(\ref{bpsi3}) is valid at {\em any\/} given point $\boldx$.
This can be accomplished only if the \fnc{}al $\Psbld$ has the structure of the continuous
product of identical \tmm{}s at all points $\boldx$, i.e. up to a normalization factor 
which includes $\varkappa$ \and{} $\sqrt{-h}$, 
\beq\label{schrod}
\Psbld \sim 
  {\sf Tr} \left \{\prod_\boldx
e^{-i\phi(\boldx)\gma^i\partial_i\phi(\boldx)/\varkappa} 
 \underline{\gma}{}_0
 \Psi_\Si (\phi(\boldx), \boldx, t)
\right \}{}_{\mbox{\Large $\rvert$} \scriptstyle
 \varkappa 
 \mapsto 
 \gma_0\dlt(\mathbf{0})/ \sqrt{-g}  } . 
\eeq 
Thus \myy{} have \ota{}ed the expression of the \Sch{} wave \fnc{}al in \tmm{}s of
\prc{} wave \fnc{}s. This formal continuous product expression can be understood 
as a multidimensional product integral \cite{prodint,slavik} 
with an invariant measure $\sqrt{-h} d\boldx$ 
(c.f. (\ref{limh})) \dnt{}d as $\prodi_\boldx f(\boldx)^{\sqrt{-h} \rd \boldx}$: 
\beq \label{print}
\Psbld \sim 
  {\sf Tr} \left \{   
  \underset{\!\!\boldx}{\scalebox{1.5}{$\displaystyle \prodi$}} 
  e^{-i\phi(\boldx)\gma^i (\boldx) \partial_i\phi(\boldx)/\varkappa}
\underline{\gma}{}_0 \Psi_\Si (\phi(\boldx), \boldx, t)
\right \}{}_{\mbox{\Large $\rvert$} \scriptstyle
 \underline{\gma}{}_0\frac1\varkappa 
 \mapsto 
 \sqrt{-h}\rd\boldx  } .  
\eeq  
This result looks remarkably similar to the result in flat \spt{} \cite{adv2}, differing only in that the Dirac matrices are now $\boldx$-dependent \and{} the spatial integration 
measure $\rd \boldx$ is replaced \by{} the invariant one.

\bigskip


\noindent 
Summarizing the results of the above consideration of (\ref{dtbpsi1}), \myy{} have shown that: 
 
\begin{itemize}

\item the potential \tmm{} $V$ reproduces the potential \tmm{} in (\ref{fuse}) 
in the \lmt{}ing \cse{} (\ref{lim});

\item in the same \lmt{}ing \cse{}, the \tmm{} $IV$ reproduces the second \fnc{}al \drv{}ive \tmm{} in the canonical \hlt{}ian operator in (\ref{fuse}) up to some  additional \tmm{}s 
in (\ref{e218});

\item the required cancellation of those additional \tmm{}s together 
with the \tmm{} $II$ in (\ref{dtbpsi1}) allows us to 
\ota{} an explicit product integral formula for the \Sch{} wave \fnc{}al in 
\tmm{}s of the \prc{} wave \fnc{}, Eq. (\ref{print}),  \and{} to reproduce the second \tmm{} 
in (\ref{fuse}) which is responsible for non-ultralocality; 

\item the \tmm{}s $I$ \and{}  $III$ do not contribute to the \fnc{}al \Sch{} \equn{} 
(\ref{fuse}) if the \fd{}s $\phi(\boldx)$ \and{} $\Psi_\Si(\phi(\boldx),\boldx,t)$ are \vsh{}ing at the spatial infinity. 

\end{itemize}

Thus, in static \spt{}s, \myy{} have demonstrated that 
the canonical 
\fnc{}al \drv{}ive \Sch{} \equn{} (\ref{fuse})  \and{} the explicit product integral formula (\ref{print}) 
relating the \Sch{} wave \fnc{}al with the \clf{}-valued 
\prc{} wave \fnc{} follow from the \prc{} \Sch{} \equn{} (\ref{crvns}) 
in the symbolic \lmt{}ing \cse{} \wnn{} 
$\underline{\gma}{}_0\varkappa$  is replaced \by{} (a regularized) invariant delta-\fnc{} 
$\dlt{\bf(0)}/\sqrt{-h}$ 
at equal spatial points, 
i.e. essentially, \by{} the UV cutoff of the total volume of the momentum space.

\section{Conclusion}

We  investigated a  \con{} \bewn{} the description of \qnt{}um 
scalar theory in curved \spt{} based on \prc{} \qnt{}ization 
\and{}  the st\and{}ard description using the \fnc{}al \Sch{} representation 
resulting from the canonical \qnt{}ization. 
Conceptually, there is a huge gap \bewn{} the canonical description 
in \tmm{}s of a \Sch{} wave \fnc{}al of \fd{} configurations $\phi(\boldx)$ on 
fixed-time hypersurfaces labelled \by{} $t$ \and{} the \prc{} description in \tmm{}s 
of a \prc{} wave \fnc{}, which is a section of the \clf{} bundle \cite{berline} 
over the  finite-dimensional bundle with the fiber coordinates $\phi$ \and{} the 
base coordinates $x^\mu$. 
 Nevertheless, 
\myy{} demonstrated in the \cse{} of static \spt{}s 
that the \ltr{} can be derived from the former in the 
\lmt{}ing \cse{} of an infinitesimal value of $1/ \varkappa$ 
\wnn{} the \clf{} algebra element $\underline{\gma}{}_0 / \varkappa $ can be 
replaced \by{} 
 or mapped to the \dfrt{}ial form representing an infinitesimal invariant spatial volume element 
$\sqrt{-h}\,\mathrm{d} \boldx$.  
In this (symbolic) \lmt{}ing \cse{},  
\myy{} \myy{}re able to derive the st\and{}ard \fnc{}al \drv{}e \Sch{} \equn{} for 
the \qnt{}um scalar \fd{} in static curved \spt{}s from  the \prc{} 
\Sch{} \equn{} for the same physical system, \and{} also to \ota{}, 
up to a normalization factor,  
an expression of the \Sch{} wave \fnc{}al of \qnt{}um scalar \fd{}  theory in \tmm{}s of 
a multiple product integral of \prc{} wave \fnc{}s \rsd{} to a \fd{} configuration 
 $\phi = \phi (\boldx)$ at a fixed moment of time, Eq. (\ref{print}). 
 
Our result confirms \and{} generalizes to static \spt{}s  
the statement from our previous papers \cite{adv1,adv2,rmp18}
that the st\and{}ard \fnc{}al \Sch{} representation of 
\qnt{}um \fd{} theory is a certain  (symbolic) \lmt{}ing \cse{} of the theory 
of \qnt{}um \fd{}s \ota{}ed \by{} \prc{} \qnt{}ization. 
 While the  
former, in order to be a \myy{}ll-defined theory at least on the physical level of rigour, 
is known to require an ad hoc regularization (e.g. a point-splitting  in the second variational 
\drv{}ive in the \fnc{}al \drv{}ive \Sch{} \equn{} (\ref{fuse})), which typically 
introduces a UV cutoff scale $\Lambda$ as a necessary additional element of the theory 
removed \by{} a subsequent renormalization,   
 the \prc{} formulation is ``already regularized" 
since the ultraviolet scale $\varkappa$ is an inherent element of the \prc{} \qnt{}ization procedure. 

One can still wonder if 
$\varkappa$ is a fundamental scale or 
an auxiliary element of \prc{} \qnt{}ization of \fd{}s which should 
be removed from the physical results  \by{} a procedure similar to the usual renormalization. 
The \ltr{} point of view is supported \by{} the observation that $\varkappa$ actually disappears from the observable \qnt{}ities of the theory in the \cse{} of free \fd{}s. On the other h\and{}, the fundamental nature of the scale $\varkappa$ is supported \by{} our recent estimation of the 
mass gap in the \prc{} formulation of \qnt{}um pure SU(2) gauge theory \cite{my-massgap} \and{} a naive estimation of the cosmological constant from the \prc{}ly \qnt{}ized pure Einstein gravity \cite{qg}. Surprisingly, both estimations point to a subnuclear scale of $\varkappa$, 
thus a further research is required to underst\and{} this fact. 

The consideration in this paper has also allo\myy{}d us to realize that the spin-\con{} 
matrix in the curved \spt{} generalization of \prc{} Schr\"odin\-ger \equn{}, Eq. (\ref{crvns}), acts on the \clf{}-algebra-valued \prc{} wave \fnc{} 
 via a commutator. This guarantees the Leibniz property \wnn{} the \crsp{}ing covariant \drv{}ive 
acts on the \clf{} product of two \clf{}-algebra-valued \qnt{}ities, \and{} this property 
is crucial for 
the proof in Eq. (\ref{first}) of the \vsh{}ing contribution 
of the \tmm{}s $I$ \and{} $III$ in (\ref{dtbpsi1}) to the \fnc{}al \Sch{} \equn{} (\ref{fuse}). 
This observation \imls{} that some details of the earlier demonstrations of the Ehrenfest 
theorem in curved \spt{} \cite{ehr1}  \and{} \prc{} \qnt{}um gravity \cite{ehr2} will have to be modified, which \myy{} hope to address in a future paper.

 \subsection*{\normalsize Acknowledgments}
  
I gratefully appreciate the hospitality of the School of Physics and Astronomy 
of the University of St Andrews, Scotland, and 24/7 availability of its facilities for 
research.  I also thank S. Capozziello, F. Finster, E. Minguzzi \and{} V. Zatloukal for 
useful  remarks.


{\footnotesize 

\begin{thebibliography}{99}


\bibitem{fr} K. Freese, C.T. Hill and M.T. Mueller, 
Covariant functional Schr\"odinger formalism and application to the Hawking effect, 
{\em Nucl. Phys.} {\bf B255} (1985), 693.

\bibitem{pi} S.-Y. Pi, Quantum field theory in flat Robertson-Walker space-time: functional 
Schr\"odinger picture, in {\sl Field Theory and Particle Physics}, eds.  
O. \'Eboli, M. Gomes, and A. Santoro (World Scientific, Singapore, 1990) 144-195. 

\bibitem{se1} D.V. Long and G.~M. Shore, 
The Schr\"odinger wave functional and vacuum states in curved spacetime, 
{\em Nucl. Phys. B} {\bf 530} (1998), 247 
 {\tt [arXiv:hep-th/9605004]}. 

\bibitem{se2} D.V. Long and G.~M. Shore, 
The Schr\"odinger wave function{}al and vacuum states in curved spacetime II: Boundaries and foliations, {\em Nucl. Phys. B} {\bf 530} (1998), 279 
{\tt [arXiv:gr-qc/9607032]}.

\bibitem{cr1} A. Corichi and H. Quevedo, Schr\"odinger
 representation for a scalar field on curved spacetime,	
	{\em Phys. Rev.} {\bf D66} (2002),  085025 
	{\tt [arXiv:gr-qc/0207088]}.
	
\bibitem{cr2} A. Corichi, J. Cortez and H. Quevedo,	Schr\"odinger
 and Fock representation for a field theory on curved spacetime,
{\em Ann. Phys.} {\bf 113} (2004), 446 
{\tt [arXiv:hep-th/0202070]}. 
	
\bibitem{kr} C. Kiefer, 
Quantum gravitational corrections to the function{}al Schr\"odinger \equn{}, 
{\em  Phys. Rev.} {\bf D44} (1991), 1067 

\bibitem{htf}  
B. Hatfield,  
{\sl Quantum Field Theory of Point 
Particles and Strings}  
(Reading MA, Addison-Wesley, 1992). 
S.S.~Schweber,  
{\sl An Introduction to Relativistic Quantum Field Theory}  
(Harper $\&$ Row, New York, 1961). 

\bibitem{jw} 
R. Jackiw, 
Analysis on infinite dimensional manifolds: \Sch{} representation for quantized fields, 
in {\sl Field Theory and Particle Physics}, edited by O. \'Eboli, M. Gomes, and A. Santoro 
(World Scientific, Singapore 1990) 78-143. 
 
 \bibitem{ehr1}  I.V.~Kanatchikov, 
 Ehrenfest theorem in precanonical quantization, 
 {\em J. Geom. Symmetry Phys.} {\bf 37} (2015), 43   
 {\tt [arXiv:1501.00480]}.  

 \bibitem{ehr2} I.V.~Kanatchikov, 	
{Ehrenfest theorem in precanonical quantization of fields and gravity, } 
in {\em The Fourteenth Marcel Grossmann Meeting on Recent Developments in Theoretical and Experimental General Relativity, Astrophysics, and Relativistic \fd{} Theories, Part C}, 
 edited by M. Bianchi, R.T. Jantzen, and R. Ruffini (World Scientific, Singapore, 2018)  
 2828-35, 
{\tt arXiv:1602.01083}. 


\bibitem{ivko} V.~Fock  and D.~Iwanenko, G\'eometrie quantique lin\'eaire et d\'eplacement parall\'ele, {\em Compt. Rend. Acad. Sci. Paris}, {\bf 188} (1929), 1470. 

\bibitem{plk} M.D.~Pollock, On the Dirac \equn{} in curved \spt{}, 
{\em Acta Phys. Pol.}  {\bf B41} (2010), 1827. 

 \bibitem{qs96} I.V.~Kanatchikov, 
{ Towards the Born-Weyl quantization of fields, }
{\em Int. J. Theor. Phys.} {\bf 37}  (1998), 333 
{\tt [arXiv:quant-ph/9712058]}. 
See also an earlier attempt in 
I. Kanatchikov, From the Poincar\'e-Cartan form to a Gerstenhaber algebra 
of Poisson brackets in field theory,  {\tt arXiv:hep-th/9511039}. 



 \bibitem{bl}  I.V.~Kanatchikov,  
{De Donder-Weyl theory and a hypercomplex 
extension of quantum mechanics to field theory, } 
 {\em Rep. Math. Phys.} {\bf 43} (1999), 157 
{\tt [arXiv:hep-th/9810165]}.   

\bibitem{ldz} I.V.~Kanatchikov,  
On quantization of field theories in polymomentum variables,     
{\em AIP Conf. Proc.}  {\bf 453} (1998), 356 
  {\tt [arXiv:hep-th/9811016]}. 

\bibitem{my-pla} 
 I.V.~Kanatchikov, 
 {\prc{} quantization and the Schr\"o\-dinger wave functional},
{\em Phys. Lett.} {\bf A283} (2001), 25 
{\tt [arXiv:hep-th/0012084]}.

\bibitem{my-ym1} I.V.~Kanatchikov, 
\prc{} quantization of Yang-Mills fields and the functional Schr\"o\-dinger
 representation,
{\em Rep. Math. Phys.} {\bf 53} (2004), 181 
{\tt  [arXiv:hep-th/0301001]}. 

 \bibitem{geom-q} I.V.~Kanatchikov, 
Geometric (pre)quantization in the polysymplectic approach to field theory, 
in {\em Differential Geometry and its Applications}, eds. Krupka, O. Kowalski,
O. Krupkova and J. Slov\'ak (Worlds Scientific, Singapore, 2008) 309-322, 
{\tt  [arXiv:hep-th/0112263]}. 
I.V. Kanatchikov, Covariant geometric prequantization of fields, 
{\tt arXiv:gr-qc/0012038}
  
 \bibitem{adv1} 
I.V.~Kanatchikov, 
{ \prc{} quantization and the Schr\"o\-dinger wave function{}al revisited}, 
{\em Adv. Theor. Math. Phys.} {\bf 18}  (2014), 1249 
{\tt [arXiv:1112.5801]}.  

\bibitem{adv2}
 I.V.~Kanatchikov, 
{  On the precanonical structure of the Schr\"o\-dinger wave functional},  
{\em Adv. Theor. Math. Phys.} {\bf 20} (2016), 1377 
 {\tt [arXiv:1312.4518]}. 
 
 
 
\bibitem{rmp18} I.V.~Kanatchikov,  
Schr\"odinger wave functional in quantum Yang-Mills theory from \prc{} quantization, 
{\em Rep. Math. Phys.} {\bf 82} (2018) 373,   
{\tt [arXiv:1805.05279]}. 



\bibitem{dd} 
Th. De Donder, 
{\em Th\'eorie Invariantive du Calcul des Variations, }  
 (Gauthier-Villars, Paris, 1935).   

\bibitem{we} 
H. Weyl, 
 {Geodesic \fd{}s in the calculus of variations, } 
 {\em Ann. Math. (2)} {\bf 36} (1935), 607. 

\bibitem{ru} 
H. Rund,   
{\em The Hamilton-Jacobi Theory in the Calculus of 
Variations} (D. van Nostrand, Toronto, 1966).  

\bibitem{ka} 
H. Kastrup, Canonical theories of Lagrangian dynamical systems in physics, 
{\em Phys. Rep.} {\bf 101} (1983), 1. 

 
 \bibitem{pg1}
 I.V.~Kanatchikov, 
Canonical structure of classical \fd{} theory in the polymomentum phase space,    
{\em Rep.~Math.~Phys.} {\bf 41} (1998), 49 
{\tt  [arXiv:hep-th/9709229]};   
See also I.V. Kanatchikov, Basic structures of the covariant canonical formalism for \fd{}s based on the De Donder--Weyl theory, {\tt arXiv:hep-th/9410238}.  


  \bibitem{pg2}
 I.V.~Kanatchikov, 
 On \fd{} theoretic generalizations of a Poisson algebra,   
{\em Rep. Math. Phys.} {\bf 40} (1997),  225 
{\tt  [arXiv:hep-th/9710069]}. 

\bibitem{joseph1} F. H\'elein and J. Kouneiher, 
The notion of observable in the covariant Hamiltonian formalism for 
the calculus of variations with several variables, 
{\em Adv. Theor. Math. Phys.} {\bf 8} (2004), 735 
{\tt  [arXiv:math-ph/0401047]}. 

\bibitem{joseph2}
F. H\'elein and J. Kouneiher, 
Covariant Hamiltonian formalism for the calculus of variations with several variables, 
{\em Adv. Theor. Math. Phys.} {\bf 8} (2004), 565 
{\tt  [arXiv:math-ph/0211046]}.

\bibitem{my-loday} I.V.~Kanatchikov, 
Novel algebraic structures from the polysymplectic form in \fd{} theory, 
in  {\sl GROUP21, Physical Applications and Mathematical Aspects of Geometry, Groups and Algebras,} vol. 2, eds. H.-D. Doebner, P. Nat\tmm{}ann , W. Scherer and C Schulte 
(World Scientific, Singapore, 1997) 894, 
 {\tt [arXiv:hep-th/9612255]}. 

\bibitem{romer} M. Forger, C. Paufler and H. R\"omer, 
The Poisson bracket for Poisson forms in multisymplectic \fd{} theory, 
  {\em Rev. Math. Phys.} {\bf 15} (2003), 705 	
  {\tt  [arXiv:math-ph/0202043]}. 
  
  \bibitem{my-dkp} I.V.~Kanatchikov, 
On the Duffin-Kemmer-Petiau formulation of the covariant Hamiltonian dynamics in \fd{} theory, 
{\em Rep. Math. Phys.} {\bf 46} (2000), 107 
{\tt  [arXiv:hep-th/9911175]}. 

 
 \bibitem{my-dirac} I. Kanatchikov,
{On a generalization of the Dirac bracket in the De Donder-Weyl Hamiltonian formalism}, 
in {\em Differential Geometry and its Applications},  
eds. Kowalski O., Krupka D., Krupkov\'a O., and Slov\'ak J. 
 (World Scientific, Singapore, 2008) 615-625, 
{\tt [arXiv:0807.3127]}.   
 

\bibitem{ijtp2001} I.V. Kanatchikov, 
Precanonical quantum gravity: quantization without 
the space-time decomposition, 
{\em Int. J. Theor. Phys.\/} {\bf 40} (2001), 1121 
{\tt arXiv:gr-qc/0012074}.  
See also: I.V. Kanatchikov, From the De Donder–Weyl Hamiltonian formalism to quantization
of gravity, {\tt arXiv:gr-qc/9810076}; 
I.V. Kanatchikov, Quantization of gravity: yet another way, 
{\tt arXiv:gr-qc/9912094}; 
I.V. Kanatchikov, Precanonical perspective in quantum gravity, {\em Nucl. Phys. Proc. Suppl.}  {\bf88} (2000) 326, {\tt arXiv:gr-qc/0004066}.




\bibitem{qg} 
Igor V. Kanatchikov, 
 {On the ``spin-connection foam" picture of quantum gravity  
 from precanonical quantization, }  
 in {\em The Fourteenth Marcel Grossmann Meeting on Recent Developments in Theoretical and Experimental General Relativity, Astrophysics, and Relativistic \fd{} Theories, Part D}, 
 edited by M. Bianchi, R.T. Jantzen, and R. Ruffini (World Scientific, Singapore, 2018) 
 3907, 
{\tt [arXiv:1512.09137]}.  



I.V. Kanatchikov, 
{On \prc{} quantization of  gravity in spin \con{} variables}, 
{\em AIP Conf. Proc. } {\bf 1514} (2012), 73 
 {\tt [arXiv:1212.6963]}. 

I.V. Kanatchikov, 
{ De Donder-Weyl Hamiltonian formulation and \prc{} quantization 
of  vielbein gravity}, 
{\em J. Phys. Conf. Ser.} {\bf 442}  (2013), 012041 
{\tt [arXiv:1302.2610]}. 

I.V. Kanatchikov, 
{On \prc{} quantization of  gravity}, 
 {\em Nonlin. Phenom. Complex Sys. (NPCS)} 
 {\bf 17} (2014), 372 
{\tt [arXiv:1407.3101]}.
 
 


 \bibitem{my-massgap} Igor V. Kanatchikov, 
 On the spectrum of \dwh{} of quantum SU(2) gauge \fd{}, 
 {\em Int.~J.~Geom.~Meth.~Mod.~Phys.} {\bf 14} (2017), 1750123 
 {\tt [arXiv:1706.01766]}.
 
\bibitem{new38} M.E. Pietrzyk and I.V. Kanatchikov, 
On the polysymplectic integrator for the short pulse equation{}, 
 {\tt arXiv:1512.09105};  
 M. Pietrzyk, I. Kanatt\v{s}ikov and A. Demircan, 
 On the compression of ultrashort optical pulses
beyond the slowly varying envelope approximation, 
in {\em 2008 International Conference on Numerical Simulation of Optoelectronic Devices (NUSOD)},  eds. J. Piprek and E. Larkins (IEEE, 2008) 59;  
M. Pietrzyk and I. Kanatt\v{s}ikov, 
Multisymplectic analysis of the short pulse \equn{},
WIAS preprint No. 1278 
(WIAS, Berlin, 2007). 

 \bibitem{new39}  M.E. Pietrzyk, 
 Polysymplectic integrator for the short pulse \equn{} and numerical general relativity, 
 in {\em The Fourteenth Marcel Grossmann Meeting on Recent Developments in Theoretical 
 and Experimental General Relativity, Astrophysics, and Relativistic \fd{} Theories, Part C}, 
 eds. M. Bianchi, R.T. Jantzen, and R. Ruffini (World Scientific, Singapore, 2018)
 2677. 
 
 \bibitem{new40} I.V. Kanatchikov, 
 Precanonical structure of the Schr\"odinger wave function{}al  in curved space-time, 
 work in progress. 
 
  
\bibitem{prodint} 
V.~Volterra and B.~Hostinsk\'y, 
 {\it Op\'erations Infinit\'esimales Lin\'eaires}  
 (Gauthier-Villars, Paris, 1938). 
 
 \bibitem{slavik} 
A.~Slav\'\i k,  {\it Product Integration, its History and Applications} 
 (Matfyzpress, Prague, 2007). 
 \url{http://www.karlin.mff.cuni.cz/~slavik/product/product_integration.pdf}


\bibitem{saunders} D.J. Saunders, {\em The Geometry of Jet Bundles}  
(Cambridge University Press, Cambridge, 1989). 

\bibitem{olver}
P.J. Olver, {\em Applications of Lie Groups to Differential \equn{}s} 
(Springer-Verlag, NY, 1986). 

\bibitem{klauder} J.R. Klauder, {\em Beyond Conventional Quantization} 
(Cambridge University Press, Cambridge, 2000).  
 J.R. Klauder, Ultralocal scalar \fd{} models, 
{\em Commun. Math. Phys.} {\bf 18} (1970) 307. 

\bibitem{berline} N. Berline, E. Getzler, M. Vergne, {\em  Heat Kernels and Dirac Operators},   
(Springer-Verlag, New York, 1992). 



\end{thebibliography}

 }
\end{document}